\documentclass[a4paper,12pt]{article}
\usepackage{epsfig}
\usepackage{graphicx}
\usepackage{subfigure}
\begin{document}

\title{\textbf{Dynamics of $\varphi^{4}$ Kinks on a Class of Space Dependent Potentials}}
\author{Jassem H. Al-Alawi\thanks{e-mail address:alawi\_j@jic.edu.sa}
\\ Department of General Studies,Jubail Industrial College, \\
 Jubail  31961, Saudi Arabia\\
}

\date{\today}

\maketitle

\begin{abstract}
In this paper we study the dynamics of $\varphi^{4}$ kinks, generated by considering a particular argument of $\varphi^{4}$ field, in the presence of class of smooth space dependent potentials \emph{ie}. barriers and wells. Various type of these potentials are produced via two-parameter family. These parameters control the shapes, widths and heights of such potentials. The dynamics are presented through the plotting of the trajectories. We compare numerical calculated critical velocities to theoretical calculations.
\end{abstract}

\section{Introduction}
Scattering of topological solitons that admit kink ( anti-kink) solutions in presence of obstructions ie. barriers/wells have been given a considerable attention in some recent research work[5-10]. Fei, et al [3] was first to investigate the interaction of soliton with a potential obstacle and many of the results in [3] were explained in [4]. These studies have shown that solitons behave like a point particle. They move in a well defined trajectories. When they meet a barrier they slows down and with enough energy they can come over the barrier and get transmitted otherwise they get reflected back. In case of a potential well, they behave also similar to a point particle, in the sense they speed up in the well. However, solitons arise from classical field theory are expected to behave as a classical point particle but they don't. Point particles are always transmitted but solitons can be trapped in the well and can be reflected by the well which can be considered as peculiar behavior. A considerable amount of radiation is generated as a result of this peculiarity behavior [1-2,11].

 The obstruction are constructed through the coupling constant $\lambda$ which is made to be zero faraway from the obstruction and is required to have a non zero value in a certain region of space and so these potentials are square barriers and square wells. The obstructions can be made space dependent i.e. $\lambda\left(x\right)$ but with no exact solitonic solutions. These space dependent potentials are considered to be a perturbation in the theory.

 In this paper following [5] we look for a class of space dependent potentials which are smooth and pose analytic solution for a static kink located at the center of these obstructions \emph{ie}. $x=0$. And so in this paper the parameter $\lambda$ is a function of space \emph{ie}. $\lambda\left(x\right)$. This parameter is constrained to have a value of $1$ as $\vert x\vert\rightarrow \infty$ and with this constraint the kink(anti kink)solution reduces to the usual $\varphi^{4}$ solution. However, close to or at the obstruction, $\vert x\vert \rightarrow 0$ new soliton solution is developed. In this paper our mathematical approach to developing such two different solutions one at the obstruction and the other one faraway from the obstruction is more generic.

 In section 2 of this paper we review some basic facts of $\varphi^{4}$ kink. In [5] they have found a two-parameter family of a class of smooth potentials for sine-Gordon kinks located at $x=0$. In section 3 of this paper we explain our mathematical approach to construct as is the case in [5] a two-parameter family of class of smooth potentials of $\varphi^{4}$ kinks located at $x=0$. This has been achieved through some constraints imposed over what we call an argument function of the field $g\left(x\right)$. This argument field function has been selected carefully so to allow kink (anti kink )solution in the $\varphi^{4}$ model. The choice of the argument function determines the shape of these potentials and this in turn affects the dynamics of the kinks when they are in these potentials and when they come out.


  In section 4 we studied the dynamics of scattering of class of solitons with $n=1,3,5,\dots$ solutions with the corresponding  potential barriers and potential wells. Trajectories, $x\left(t\right)$, of all known phenomena were plotted. Definitely a different choice of $g\left(x\right)$ will result in a different trajectories. In section 5 we calculated the critical velocities of a kink interacting with a barrier using the moduli space approximation and compared the results with numerical calculations.

\section{\bf{Review of $\lambda\varphi^{4}$ Kink}}

A kink solution for $\varphi^{4}$ field theory in (1+1) dimensions is one of the simplest model that admits kink solution. The model is described by
the lagrangian density
\begin{equation}
\mathcal{L}=\frac{1}{2}\partial_{\mu}\varphi\partial^{\mu}\varphi-\lambda\left(\varphi^{2}-1\right)^{2},  \qquad \mu=0,1
\end{equation}

where $\varphi\left(x, t\right)$ is a scalar field and $\lambda$ is a coupling constant. Applying the Euler-Lagrange equation leads to the following field equation

\begin{equation}
\varphi_{tt}-\varphi_{xx}+4\lambda\varphi\left(\varphi^{2}-1\right)=0.
\end{equation}


The stationary soliton solution for the static field $\varphi=\varphi\left(\,x\right)$, for which $\lambda$ is constant, can be obtained by solving the static field equation

\begin{equation}\label{static eq}
\varphi_{xx}-4\lambda\varphi\left(\varphi^{2}-1\right)=0.
\end{equation}

The static field that solves (\ref{static eq}) is

\begin{equation}\label{wojtek}
\varphi\left(x\right)=\pm\tanh\left(\sqrt{\lambda}\left(x-x_{0}\right)\right),
\end{equation}

where$\pm$ corresponds to kink(anti kink) solutions and $x_{0}$ is a constant of integration which corresponds to the location of the soliton.

The total energy of static $\varphi^{4}$ kink is given by

\begin{equation}
E=\int_{-\infty}^{\infty}dx \left(\frac{1}{2}\left(\frac{d\varphi}{dx}\right)^{2}+\lambda\left(\varphi^{2}-1\right)^{2}\right).
\end{equation}
\\
The energy can be evaluated to be $\frac{4}{3\sqrt{\lambda}}$ and so the Bogomolny bound is
\begin{equation}
E\geq \frac{4}{3\sqrt{\lambda}}.
\end{equation}

The field and energy solutions are singular as $\lambda\rightarrow 0$ which illustrate a general feature of solitons.

The theory is Lorentz invariant and so the solution (\ref{wojtek}) can be boosted to obtain a time-dependent solution thus the field is allowed to evolve with  time

\begin{equation}
\varphi\left(x,t\right)=\pm\tanh\left(\sqrt{\lambda}\gamma\left(x-x_{0}-ut\right)\right),
\end{equation}

 where $u$ is the velocity of the soliton and $\gamma=\frac{1}{\sqrt{1-u^{2}}}$ is the Lorentz factor.

\section{\bf{A Class of Space Dependent Potentials}}

In this paper we are interested in the parameter $\lambda$ as a function of space \emph{ie}. $\lambda\left(x\right)$ and so we are looking for kink(anti kink) solution that can solve the following static field equation

\begin{equation}\label{static}
\varphi_{xx}-4\lambda\left(x\right)\varphi\left(\varphi^{2}-1\right)=0.
 \end{equation}

To solve the static field equation (\ref{static}) with the parameter $\lambda$ is space dependent we will consider a solution of the form

\begin{equation}\label{*}
\varphi\left(x\right)=\tanh\left(g\left(x\right)\right).
\end{equation}

where $g\left(x\right)$ is the field argument function which we are going to write it down explicitly after imposing some constraints on it so that we can have a possible static kink ( anti kink) solution.  The range of (\ref{*}) is
\begin{equation}
-1 \leq \tanh\left(g\left(x\right)\right)\leq 1 ,
\end{equation}

and so the range of  $g\left(x\right)$ is $\left(-\infty,\infty\right)$.

Inserting the static field (\ref{*}) into the static filed equation (\ref{static})with $\lambda$ now is made space dependent gives

\begin{equation}\label{**}
\frac{d^2g}{d x^2}-2\left(\frac{d g}{dx}\right)^2 \tanh\left(g\left(x\right)\right)+4 \lambda\left(x\right)\tanh\left(g\left(x\right)\right)=0.
\end{equation}

 With the parameter $\lambda$ is being space dependent we can impose some constraints on the argument function $g\left(x\right)$. From (\ref{**}) $\lambda\left(x\right)$ is

 \begin{equation}\label{***}
 \lambda\left(x\right)=\frac{1}{2}\left(g'^{2}\left(x\right)-\frac{g''\left(x\right)}{2\tanh\left(g\left(x\right)\right)}\right)
 \end{equation}

 where $g'\left(x\right)=\frac{dg\left(x\right)}{dx}$ and  $g''\left(x\right)=\frac{d^{2}g\left(x\right)}{dx^{2}}$. And in order to facilitate our calculations we will use$\tilde\lambda\left(x\right)$ where $\tilde\lambda\left(x\right)=2\lambda\left(x\right)$.
we require that $\tilde\lambda\left(x\right)\rightarrow 1$ and $g\left(x\right)\rightarrow x$ as $\vert x\vert \rightarrow \infty$. So, when the soliton is faraway from the obstruction the kink solution reduces to (\ref{wojtek}) since the parameter $\lambda$ is constant. That is we want the usual $\lambda\varphi^{4}$ kink (\ref{wojtek}) to be the asymptotic solution to the field equation (8). We can see from(\ref{***}) that if $
g\left(x\right)=x$ which is the usual argument for the kink field then $\tilde\lambda\left(x\right) =1$. Also because the soliton field of this model is centered at the origin we will demand that $g\left(x\right)\rightarrow 0$ as $x\rightarrow 0$.

In order to obtain a kink (anti kink) solution we need the function $g\left(x\right)$ to be monotonic. So, We further set another constraint on the function $g\left(x\right)$ requiring that the

\begin{center}

$g'\left(x\right)>0$, \qquad  kink solution \\

$ g'\left(x\right)<0$, \qquad anti kink solution

\end{center}

Kink solution corresponds to an increasing monotonic function and anti kink solution corresponds to a monotonic decreasing function.

We further put some constraints on our selection of the function $g\left(x\right)$. These constraints arise from the constraints that we have already mentioned for $\tilde\lambda\left(x\right)$.

And so, the requirement that the limit as $\vert x\vert\rightarrow \infty$, $\tilde\lambda\left(x\right)\rightarrow 1$ implies that

\begin{equation}\label{1}
\lim_{\vert x\vert \to \infty} g'\left(x\right)=1,
\end{equation}

\begin{equation}\label{2}
\lim_{\vert x\vert\to \infty}g''\left(x\right)=0.
\end{equation}

So as to have smooth potentials we require that the function $g\left(x\right)$ to be differentiable and continuous on $x\in \mathbf{R}$. The parameter $\lambda\left(x\right)$ must satisfy the differentiability and continuity conditions for every $x\in \mathbf{R}$ and so when expanded using Taylor's expansion we obtain

\begin{equation}
\tilde\lambda\left(x\right)=\tilde\lambda\left(0\right)+x\tilde\lambda'\left(x\right)\vert_{x=0}+\frac{x^{2}}{2}\tilde\lambda''\left(x\right)\vert_{x=0}+\dots.
\end{equation}

Differentiating (\ref{***}) we get

\begin{equation}
\tilde\lambda'\left(x\right)=2g'\left(x\right)g''\left(x\right)+\frac{1}{\sinh\left(g\left(x\right)\right)^{2}}g'\left(x\right)g''\left(x\right)-\coth\left(x\right)g'''\left(x\right),
\end{equation}

and using the constraints we have set we obtain

 \begin{equation}
 \lim_{x\to 0}\tilde\lambda'\left(x\right)=0.
 \end{equation}

Therefore, the only none zero term is the zeroth order i.e. $\tilde\lambda\left(x\right)=\tilde\lambda\left(0\right)$. We will consider only the obstructions located at $x=0$.

The energy density is given by

\begin{equation}
\epsilon\left(x\right)=\left(g'^{2}\left(x\right)+\tilde\lambda\left(x\right)\right)\frac{1}{2\cosh\left(g\left(x\right)\right)^{4}}.
\end{equation}

On the limit as $\vert x\vert\rightarrow \infty$ the energy density reduces to

\begin{equation}
\epsilon\left(x\right)= \frac{1}{\cosh\left(x\right)^{4}}.
\end{equation}

The total energy of the static field \emph{ie}. rest mass energy on the asymptotic limit is given by

\begin{equation}
E=\int_{-\infty}^{\infty }dx \frac{1}{\cosh\left(x\right)^{4}}=\frac{4}{3}.
\end{equation}

We are now to select carefully an argument function amongst many possibilities.The choice of the argument function determines the shape of these potentials and this in turn affects the dynamics properties of the kinks when they are in these potentials and when they come out. The dynamics are contained in this choice of $g\left(x\right)$.

A possible argument function that give rise to a kink (anti kink ) field and  satisfy all the constraints that have been set is

\begin{equation}\label{g1}
g\left(x\right)=a_{1}x+b_{1}\tanh^n\left(x\right), \qquad n=1,3,5,7,\dots...,
\end{equation}

where the parameters $a_{1}"$ and $b_{1}$ have been inserted to take control over the obstructions. As $\vert x\vert\rightarrow \infty$, the term $a x$ in (\ref{g1}) will be dominated and so

\begin{center}
$a_{1}>0$ , \qquad kink solution

$a_{1}<0$ , \qquad anti kink solution
\end{center}

We will consider only the kink solution for which $a_{1}>0$.

By using (\ref{***})we can solve for the space dependent potentials i.e.$\lambda\left(x\right)$, which will from now on be designated as $\lambda_{n}\left(x\right)$

\begin{equation}
\tilde\lambda_{n}\left(x\right)=\left(a_{1}+\frac{n b_{1}\tanh^{n-1}\left(x\right)}{\cosh^{2}\left(x\right)}\right)^{2}-\frac{n b_{1} \tanh^{n}\left(x\right)\coth\left(a_{1}x+b_{1}\tanh^{n}\left(x\right)\right)}{\cosh^{2}\left(x\right)}\left(\frac{n-1}{2\sinh^{2}\left(x\right)}-1\right).
\end{equation}

Asymptotically,$\tilde\lambda_{n}\left(x\right)$ is

\begin{equation}
\lim_{\vert x \vert\to \infty}\tilde\lambda_{n}\left(x\right)=a_{1}^{2},
\end{equation}

And this sets $a_{1}=1$ because we need to satisfy the constraint that requires ,$\tilde\lambda_{n}\left(x\right)\rightarrow 1$ as $\vert x\vert\rightarrow\infty$.

Generically, with $b_{1}=0$ for $n=1,3,5,\dots.$ there will be no obstruction formed and so no interesting dynamics can be shown, see figure 1.

\begin{figure}\label{non}
\begin{center}
\includegraphics[angle=270, width=8cm]{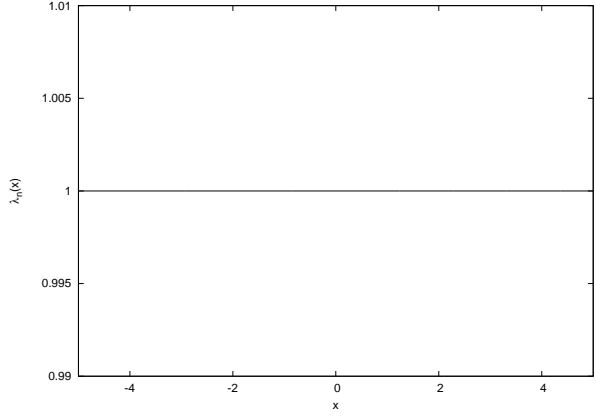}
\caption{no obstruction with $a_{1}=1,b_{1}=0$}
\end{center}
\end{figure}

We will demand for physical application that $\tilde\lambda_{n}\left(x\right)>0$.

For $n=1$

\begin{equation}
g'\left(x\right)=a_{1}+\frac{b_{1}}{\cosh^{2}\left(x\right)},
\end{equation}

and the requirement for kink solution is that $g'\left(x\right)>0$. This inequality is satisfied only when $a_{1}+b_{1}>0$, and hence $b_{1}>-1$.

\begin{equation}
\lim_{x\to 0}\tilde\lambda_{1}\left(x\right)=\frac{\left(a_{1}+b_{1}\right)^{3}+b}{a_{1}+b_{1}}.
\end{equation}

In general, we get singularities when $a_{1}+b_{1}\leq-1$

For $b_{1}\geq1$ we get a pure barriers and for $-1<b_{1}<0$ we get a pure wells.

For $n=3$ a kink solution implies that

\begin{equation}
g'\left(x\right)=a_{1}+\frac{3b_{1}\tanh^{2}\left(x\right)}{\cosh^{2}\left(x\right)}>0.
\end{equation}

This would lead to the inequality $ b_{1}>-\frac{4}{3}$. However, this is not sufficient to determine the lower bound. We require for barriers that
\begin{equation}
\lim_{x\to 0}\tilde\lambda_{3}\left(x\right)=a^{2}_{1}-\frac{3 b_{1}}{a_{1}}>0.
\end{equation}
Therefore, with $ a_{1} =1$, $b_{1}$ obeys the inequality
\begin{equation}
-\frac{4}{3}<b_{1}< 0.
\end{equation}

And for wells the requirement is
\begin{equation}
\lim_{x\to 0}\tilde\lambda_{3}\left(x\right)=a^{2}_{1}-\frac{3 b_{1}}{a_{1}}<0.
\end{equation}

 Wells, which satisfies our requirement are obtained for
 \begin{equation}
 0<b_{1}<\frac{1}{3}.
 \end{equation}

For $n=5,7,\dots$

\begin{equation}
\lim_{x\to 0}\tilde\lambda\left(x\right)=a^{2}_{1}=1.
\end{equation}

Similar to $n=3$ case, barriers are obtained for $b_{1}<0$ and wells are obtained for $b_{1}>0$. The lower bound for barriers and the upper bound for wells that satisfy our requirement can be determined numerically.

\begin{figure}\label{barriers}
\begin{center}
\includegraphics[angle=270, width=8cm]{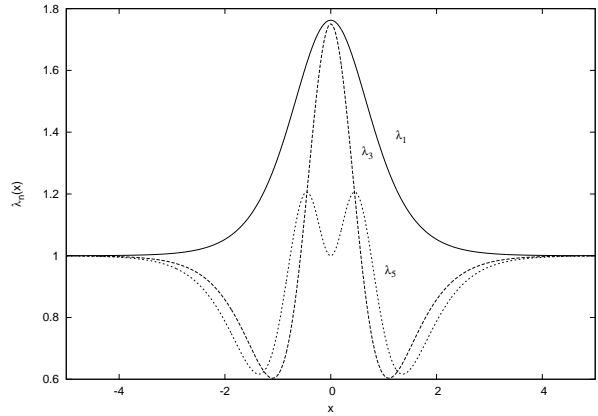}
\caption{Different types of barriers:  a pure smooth barrier, $n=1$, with ($a_{1}=1,b_{1}=0.25$), a barrier with two side narrow wells, $n=3$, with ($a_{1}=1,b_{1}=-0.25$) and a volcano barrier with two side narrow wells, $n=5$, with ($a_{1}=1, b_{1}=-0.25$)}
\end{center}
\end{figure}

In figure (\ref{barriers}), we plotted three different types of potential barriers for $n=1,3,5$ and in figure (4) we plotted three different types of wells for $n=1,3,5$.

\begin{figure}\label{wells}
\begin{center}
\includegraphics[angle=270, width=8cm]{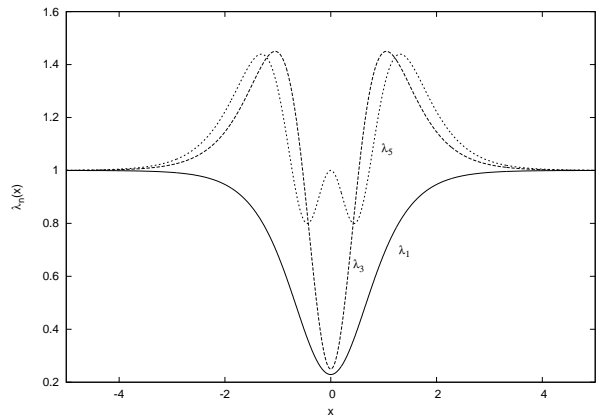}
\caption{Different types of wells: a smooth well with no humps, $n=1$ with ($a_{1}=1,b_{1}=-0.25$), a well with two side humps,$n=3$, with($a_{1}=1,b_{1}=0.25$) and a double well with two side humps, $n=5$, ($a_{1}=1,b_{1}=0.25$)}
\end{center}
\end{figure}

\begin{figure}\label{bw81}
\begin{center}
\includegraphics[angle=270, width=8cm]{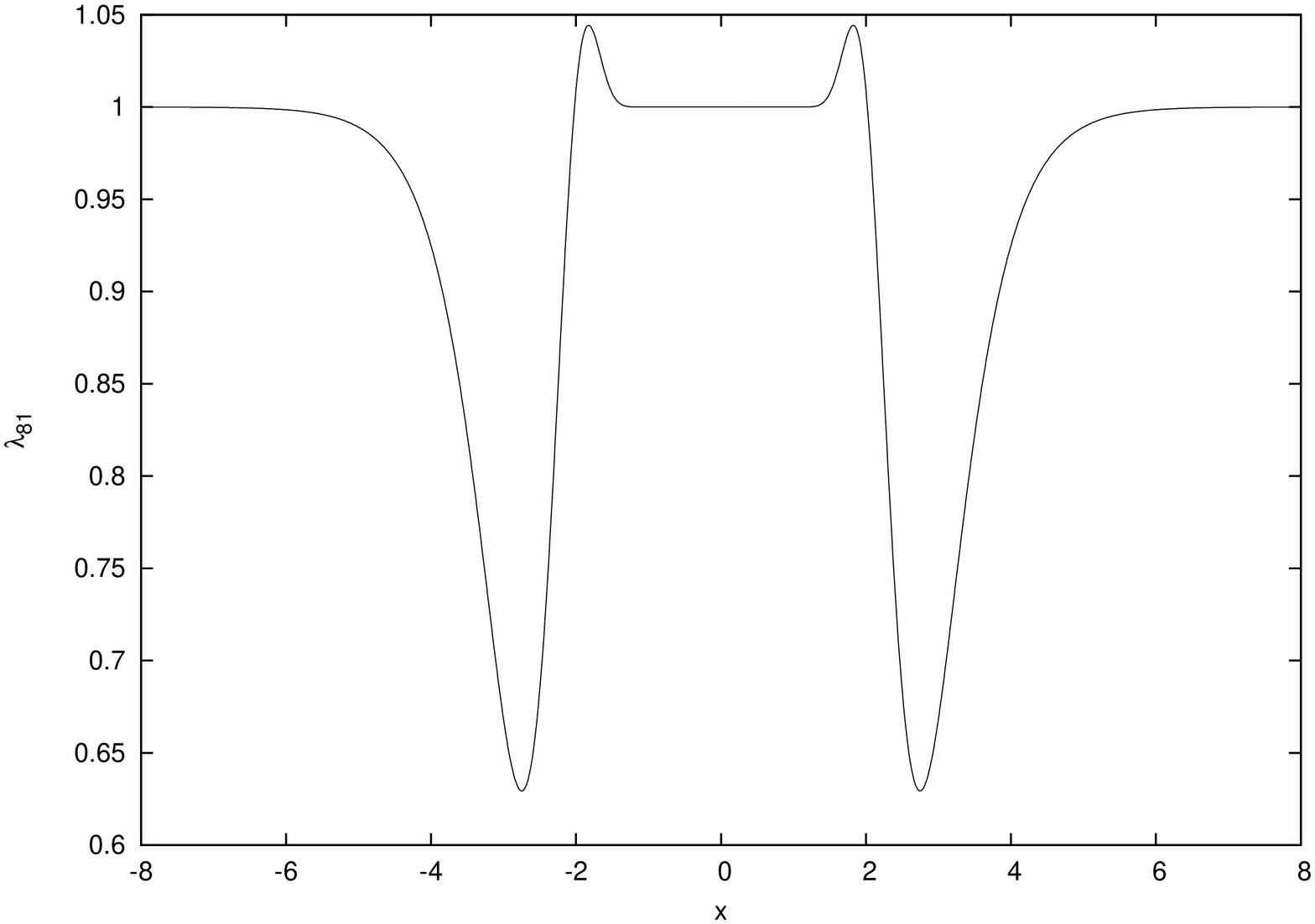}
\includegraphics[angle=270, width=8cm]{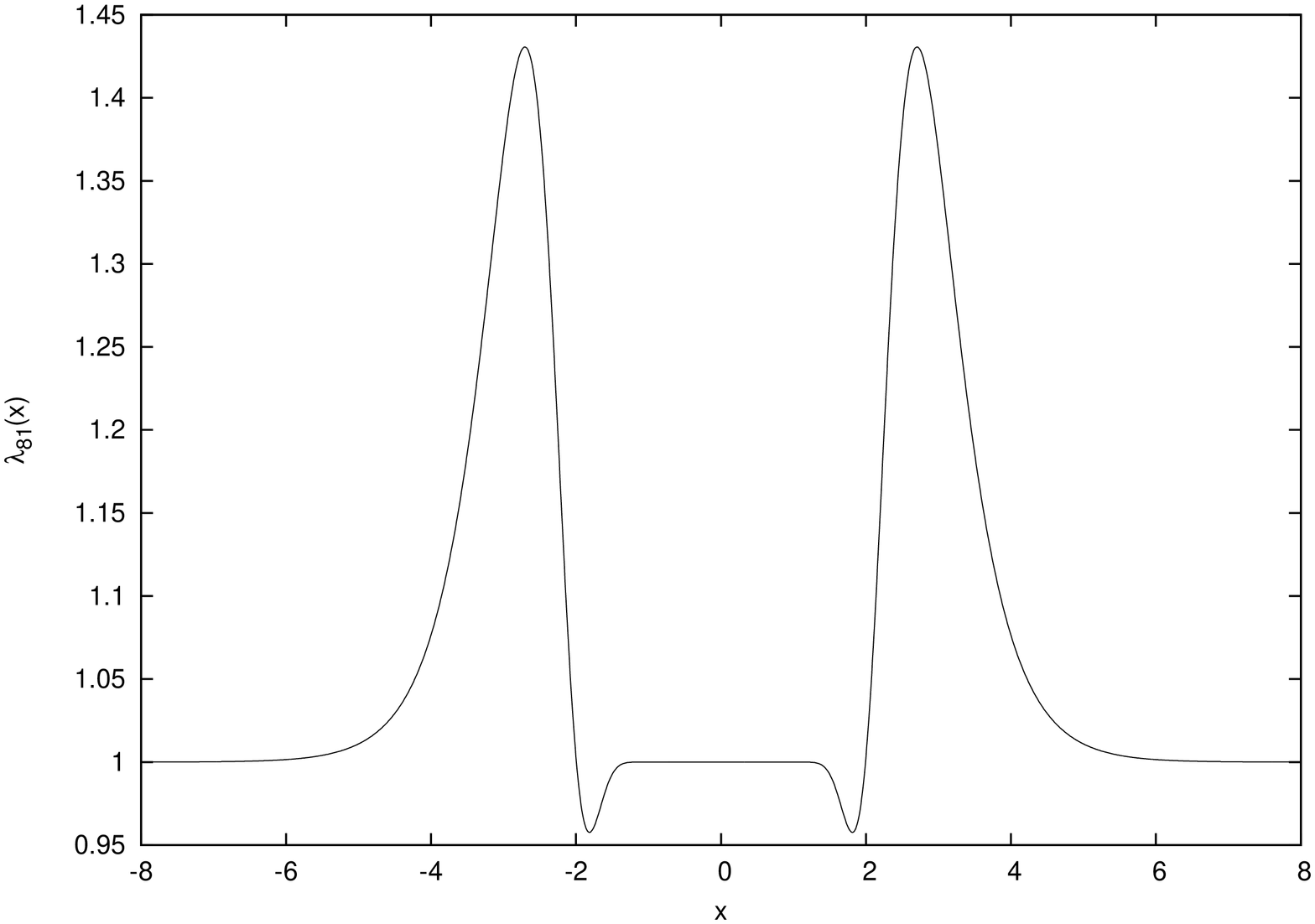}
\caption{$n=81$, a volcano barrier with two side wells and a double well with two humps}
\end{center}
\end{figure}

The barrier's top or the well's bottom are widen as the odd number increases. To illustrate this observation Figure (5) shows how the top of the barrier and the bottom's of a well are widen for $n=81$. One can see from the figures (\ref{bw81}) that the volcano barrier is almost like two separated wells and the double wells is almost like two separated barriers.

\section{\textbf{Numerical Results}}

Having we found the appropriate argument of the static field we boost the solution by performing Lorentz transformation along $x$-axis ie:

\begin{equation}
x\rightarrow\gamma\left(x-x_{0}-ut\right),
\end{equation}

where as usual $\gamma=\frac{1}{\sqrt{1-u^{2}}}$. Now, the time dependent field is given by

\begin{equation}\label{obstsol}
\varphi_{n}\left(x,t\right)=\tanh\left(\gamma\left(x-x_{0}-ut\right)+\tanh^{n}\left(\gamma\left(x-x_{0}-ut\right)\right)\right),\quad n=1,3,5,\dots
\end{equation}

The simulation of time evolution of the field were performed using 4th order Runge Kutta method.
 In this work we have used a grid containing 1201 points with lattice spacing of $dx=0.01$ and time step was chosen to be $dt=0.0025$. Hence, our lattice extends from -60 to 60 in the $x$-direction.

 Before we explore the scattering dynamics of the $\varphi^{4}$ kinks with these potentials we will compare between numerical calculations and theoretical calculations of the total energy so to give an insight on the agreement between them and to use them later in our calculations for the critical velocities.

   The total energy, $E_{n}$ is given by

\begin{equation}\label{et}
E_{n}=\int_{-\infty}^{+\infty}dx\frac{\left(g'^{2}\left(x\right)+\tilde\lambda_{n}\left(x\right)\right)}{2\cosh^{4}\left(g\left(x\right)\right)}.
\end{equation}

 Tables 1-3, show the results of the numerical calculated and theoretical calculated total energy using (\ref{et}) for $n=1,3,5$ for some various values of the parameter $b_{1}$ whereas the parameter $a_{1}=1$.

\begin{table}
\begin{center}
\begin{tabular}{|c|c|c|}
  \hline
  $b_{1}$ & $E_{1}$ (Theoretical)& $E_{1}$(Numerical) \\
  \hline
  0 & 1.333 & 1.333 \\
  0.25 & 1.712 & 1.712 \\
  0.5 & 2.064 & 2.064 \\
  0.75& 2.405 & 2.405 \\
  1 & 2.74 & 2.74 \\
  -0.25 & 0.893 & 0.893 \\
  -0.5& -0.274&-0.278\\

  \hline
\end{tabular}
\caption{soliton,s total energy for $n=1$ and $a_{1}=1$ are obtained for some values of the coefficient $b_{1}$}.
\end{center}
\end{table}

\begin{table}
\begin{center}
\begin{tabular}{|c|c|c|}
  \hline
  $b_{1}$ & $E_{3}$ (Theoretical)& $E_{3}$(Numerical) \\
  \hline
  0 & 1.333 & 1.333 \\
  0.25 & 1.169 & 1.166 \\
  0.5 & 1.015 & 1.006 \\
  0.75& 0.869 & 0.853 \\
  1 & 0.727 & 0.706 \\
  -0.25 & 1.509 & 1.515 \\
  -0.5& 1.727&1.694\\

  \hline
\end{tabular}
\caption{soliton,s total energy for $n=3$ and $a_{1}=1$ are obtained for some values of the coefficient $b_{1}$}.
\end{center}
\end{table}

\begin{table}
\begin{center}
\begin{tabular}{|c|c|c|}
  \hline
  $b_{1}$ & $E_{5}$ (Numerical)& $E_{5}$(Analytical) \\
  \hline
  0 & 1.333 & 1.333 \\
  0.25 & 1.311 & 1.303 \\
  0.5 & 1.30 & 1.273 \\
  0.75& 1.288 & 1.246 \\
  1 & 1.277 & 1.22\\
  -0.25 & 1.366 & 1.384 \\
  -0.5& 1.50&1.401\\

  \hline
\end{tabular}
\caption{soliton,s total energy for $n=5$ and $a_{1}=1$ are obtained for some values of the coefficient $b_{1}$}.
\end{center}
\end{table}

One can see from these tables that for the pure barrier and pure well which is the case with $n=1$, there is an exact agreement between the numerical and theoretical calculations. However, for the other cases with $n=3,4,5,\dots$ where the potentials posses complicated structure of a mixed barriers and wells there is no sharp agreement.

 We have explored the dynamics of soliton with $n=1,3,5$ solutions in the presence of potential barriers and potential wells. A soliton were placed faraway from an obstruction where the solution reduces to the usual $\varphi^{4}$ kink (9) and was made to move with a certain velocity toward the obstruction. As the soliton approaches the obstruction, it takes the new solution (\ref{obstsol}).  In the preceding work soliton solutions are perturbed in a certain region of space in a form of a potential barrier or a potential well. The advantage of this work is to examine the dynamics of soliton as they interact with a class of potentials which are themselves solutions of the field equation. Figures 5-14 show the field trajectories $x\left(t\right)$ for potential barriers and potential wells for $n=1,3,5$ with $a_{1}=1$.The parameter $b_{1}$ has been chosen such that for potential barriers,

 \begin{center}

$b_{1}=\cases{ 0.25 & $n=1 $ \cr
              -0.25 & $n=3,5 $ \cr} $

\end{center}

and for potential wells

 \begin{center}

$b_{1}=\cases{ -0.25 & $n=1 $ \cr
              0.25 & $n=3,5. $ \cr} $

\end{center}

As is the case in all previous work, solitons have a generic behaviour when they interact with potential barriers and potential wells if we exclude the differences of the wasted radiated amount of energy given off during the collisions with these potentials. Figures 5-14 illustrate all relevant phenomena for the kinks, $n=1,3,5$, interacting with the corresponding solutions of potential barriers and potential wells.

\begin{figure}
\begin{center}
\includegraphics[angle=270, width=8cm]{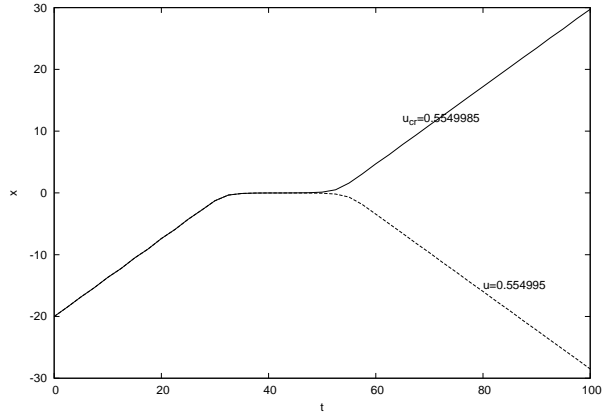}
\caption{Kink of $n=1$ elastically crosses a pure smooth barrier of $n=1$ with $b_{1}= 0.25$ with a critical velocity,  $u_{cr}=0.5549985$ and is elastically back-reflected from the same barrier with a velocity of $u=0.554995$.}
\end{center}
\end{figure}

\begin{figure}
\begin{center}
\includegraphics[angle=270, width=8cm]{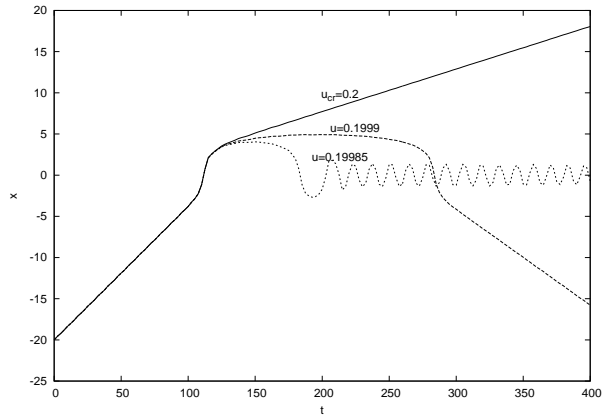}
\caption{Kink of $n=1$ inelastically crosses a pure smooth well of $n=1$ with $b_{1}=-0.25$ with a critical velocity $u_{cr} =0.2$. The kink is also inelastically back-reflected by the well $(u=0.1999)$ and becomes trapped in the well $(u=0.19985)$. }
\end{center}
\end{figure}

\begin{figure}
\begin{center}
\includegraphics[angle=270, width=8cm]{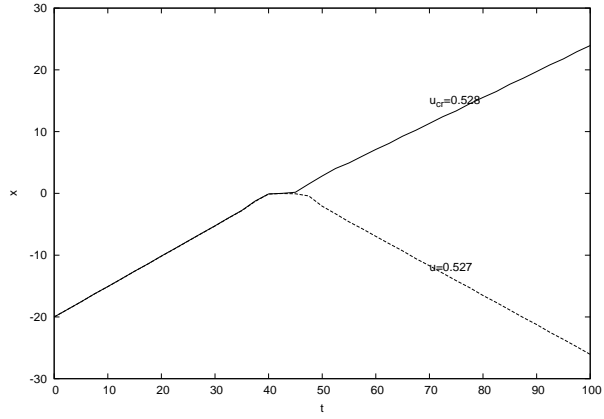}
\caption{Kink of $n=3$ inelastically overcomes a barrier with two narrow side wells, $n=1$, with $b_{1}=-0.25$ at $u_{cr} =0.528$ and is elastically back reflected from the barrier at $u=0.527$.}
\end{center}
\end{figure}

\begin{figure}
\begin{center}
\includegraphics[angle=270, width=8cm]{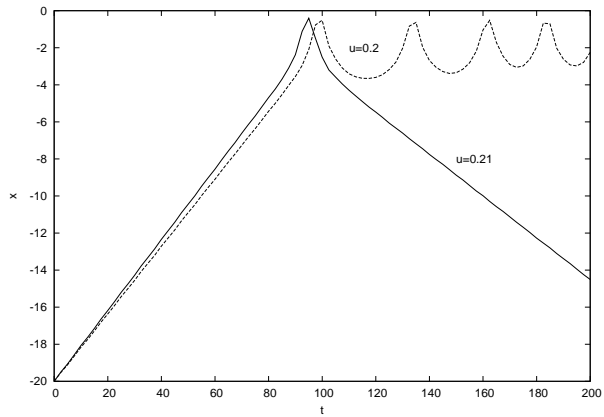}
\caption{Kink of $n=3$ is trapped in the narrow well of the barrier of $n=3$  with $b_{1}=-0.25$, at $u_{cr} =0.2$ and is back-reflected by the narrow well of the same barrier at $u=0.21$. In both cases the scattering is inelastic.}
\end{center}
\end{figure}

\begin{figure}
\begin{center}
\includegraphics[angle=270, width=8cm]{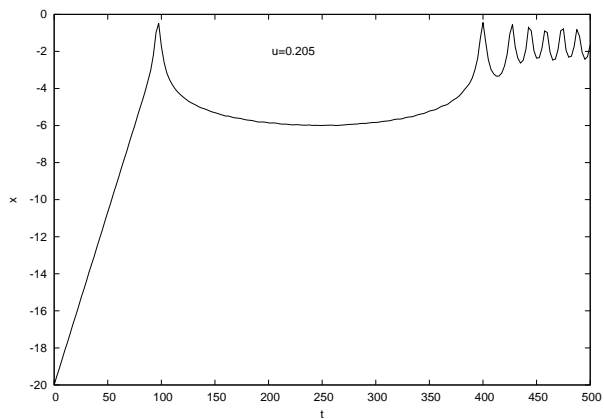}
\caption{An interesting observation of a kink of $n=3$ hardly couldn't overcome the pull of the narrow well of the barrier( $n=3$ )  with $b_{1}=-0.25$ and after a long time becomes trapped in the narrow well, $u =0.205$.}
\end{center}
\end{figure}

\begin{figure}
\begin{center}
\includegraphics[angle=270, width=8cm]{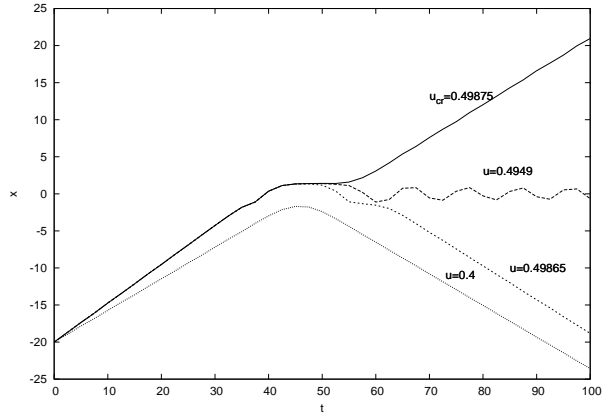}
\caption{Kink of $n=3$ is elastically back-reflected from the narrow barrier of a volcano well, $n=3$ with $b_{1}=0.25$, at $u =0.4$. The kink is inelastically crosses the volcano well at $u_{cr}=0.49875$ and is nearly elastically  back-reflected at  $u=0.498652$. It is trapped in the well, $u=0.4949$. }
\end{center}
\end{figure}

\begin{figure}
\begin{center}
\includegraphics[angle=270, width=8cm]{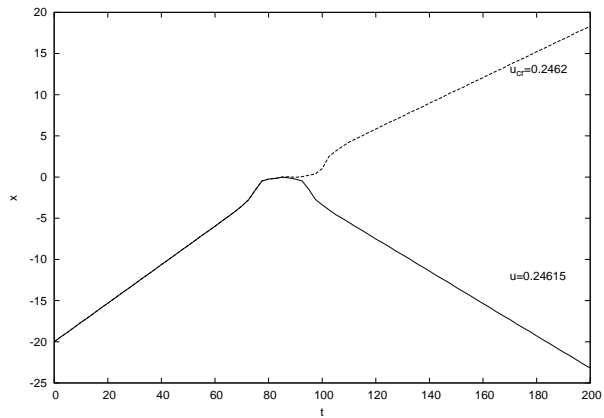}
\caption{Kink of $n=5$ inelastically crosses a volcano barrier, $n=5$ with $b_{1}=-0.25$, at $u_{cr} =0.2462$ and is nearly elastically back-reflected, $u=0.24615$.}
\end{center}
\end{figure}

\begin{figure}
\begin{center}
\includegraphics[angle=270, width=8cm]{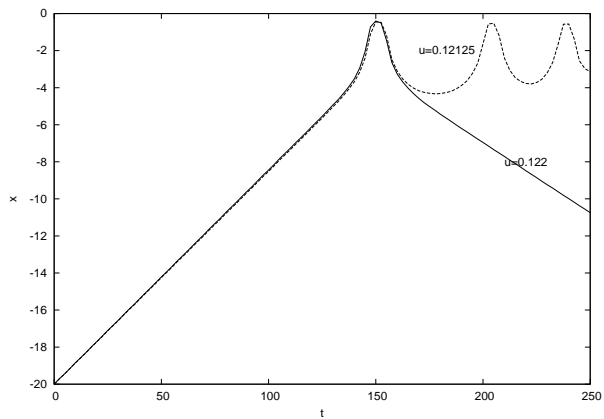}
\caption{Kink of $n=5$ is trapped in the narrow well of a volcano barrier, $n=5$,  with $b_{1}=-0.25$, at $u =0.122$  and is back-reflected by the narrow well, $u=0.12125$.}
\end{center}
\end{figure}

\begin{figure}
\begin{center}
\includegraphics[angle=270, width=8cm]{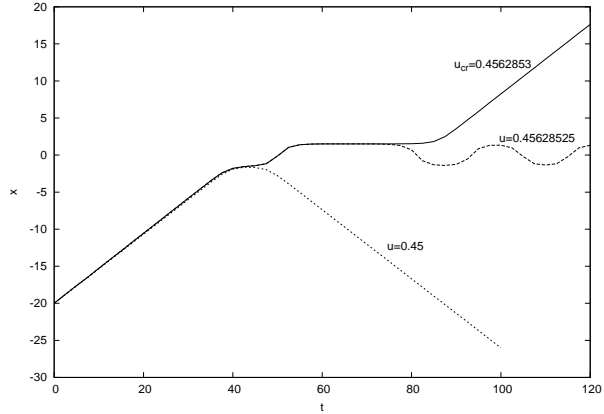}
\caption{Kink of $n=5$ nearly elastically crosses a double well with two side humps, $n=5$, with $b_{1}=0.25$, at $u_{cr} =0.4562853$ and is elastically back-reflected from the hump of the double well, $u=0.45$. It is also trapped in the double well, $u=0.45628525$ }
\end{center}
\end{figure}

\begin{figure}
\begin{center}
\includegraphics[angle=270, width=8cm]{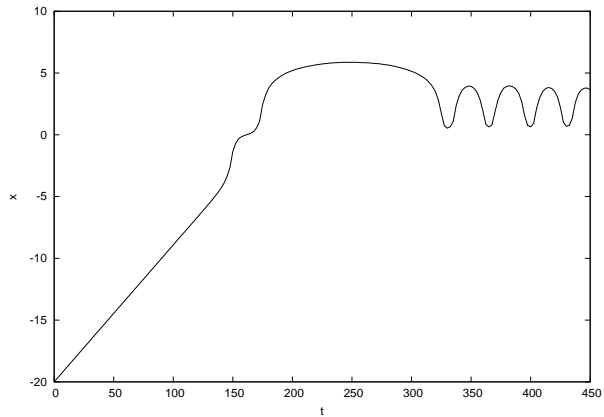}
\caption{very interesting observation of a Kink of $n=9$ is trapped in the second narrow well of a volcano barrier, $n=9$, with $b_{1}=-0.25$, at $u =0.1135$.}
\end{center}
\end{figure}

\section{\textbf{Analytical Approximation of the Critical Velocities}}

  We will calculate the critical velocities of a kink interacting with a barrier using the moduli space approximation. We make use of the ansatz

\begin{equation}\label{ansatz}
\varphi\left(x;X\right)=\tanh\left(x-X\left(t\right)\right),
\end{equation}

  where $X\left(t\right)$ is the position of the kink as a function of time. Substituting (\ref{ansatz}) into the lagrangian density (\ref{*}) we obtain

\begin{equation}
   \mathcal{L}_{n}=\frac{\left(\dot{X}^{2}-1-\tilde\lambda_{n}\left(x\right)\right)}{2\cosh^{4}\left(x-X\left(t\right)\right)}.
\end{equation}

Thus the lagrangian is

\begin{equation}\label{L}
L_{n}=\frac{1}{2}\left(\frac{4}{3}\right)\left(\dot{X}^{2}-1\right)-\frac{1}{2}\int_{-\infty}^{\infty}dx\frac{\tilde\lambda_{n}\left(x\right)}{\cosh^{4}\left(x-X\right)}.
\end{equation}

Far away from the obstruction,$x\rightarrow\infty$,so that $\tilde\lambda_{n}\left(x\right)\rightarrow 1$, the total energy is
\begin{equation}
E_{n}\left(x\rightarrow\infty\right)=\frac{1}{2}\left(\frac{4}{3}\right)\dot{X}^{2}+\frac{4}{3}.
\end{equation}

When the soliton is at rest, $\dot X=0$, the total energy is the rest mass energy, $M_{rest}$, and is

\begin{equation}
M_{rest}=\frac{4}{3}.
\end{equation}

 And when a soliton is moving toward a barrier with a critical velocity ( $\dot X=u_{cr}$), its kinetic energy at the barrier is nearly zero, $\dot X\approx 0$. Therefore, the total energy of the soliton at the barrier where, $x\rightarrow 0$, is given by

 \begin{equation}
 E_{n}\left(x\rightarrow 0\right)\approx \frac{1}{2}\left(\frac{2}{3}+\int_{-\infty}^{\infty}dx \frac{\tilde\lambda_{n}\left(x\right)}{cosh^{4}\left(x-X\left(t\right)\right)}\right).
 \end{equation}

 Energy is conserved and so

 \begin{equation}\label{energy}
 E_{n}\left(x\rightarrow \infty\right)=E_{n}\left(x\rightarrow 0\right).
 \end{equation}

 Now, we can calculate approximately the critical velocity of the soliton using(\ref{energy}).

 \begin{equation}\label{uc}
u_{cr}=\sqrt{\frac{3}{4}\int_{-\infty}^{\infty}dx  \frac{\tilde\lambda_{n}\left(x\right)}{cosh^{4}\left(x-X\left(t\right)\right)}-1}
\end{equation}
 With this approximation , the calculated critical velocities are close to the numerical ones for $n=3,5,7,9$ with exception to $n=1$, see table 4.

We can alternatively use a relativistic kinematics to calculate the critical velocities[9]. The total energy of a soliton moving with a critical velocity is given by

\begin{equation}
E_{n}=\frac{M_{rest}}{\sqrt{1-u_{cr}}},
\end{equation}

where $M_{rest}=\frac{4}{3}$. At the barrier the total energy is nearly the rest mass energy of the soliton at the top of the barrier ($M_{B}$)at $x=0$, and we have already calculated the rest mass energy at $x=0$ in tables 1,2 and 3. Conservation of energy implies that

\begin{equation}
\frac{M_{rest}}{\sqrt{1-u^{2}_{cr}}}=M_{B},
\end{equation}

 where $M_{B}$ is the rest mass energy at the top of the barrier. Thus, the critical velocity is given by

 \begin{equation}\label{rel}
 u_{cr}=\sqrt{1-(\frac{M_{rest}}{M_{B}})^{2}}.
 \end{equation}

  We present, in table (\ref{ucr}), the critical velocities calculated both numerically and theoretically using (\ref{uc}) and (\ref{rel}).

\begin{table}\label{ucr}
\begin{center}
\begin{tabular}{|c|c|c|c|c|}
  \hline
 $ n$ & $b_{1}$ &$ u_{cr}$(Theoretical) using (\ref{uc}) & $u_{cr}$(theoretical) using (\ref{rel}) & $u_{cr}$ (Numerical)\\
 \hline
  1 & 0.25 & 0.78015 & 0.627 &0.554985 \\
 \hline
  3 & -0.25& 0.525 &0.469 &0.528  \\
  \hline
  5 & -0.25 & 0.2345 & 0.218&0.2462  \\
  \hline
  7  & -0.25 & 0.1493 &0.136 &0.15  \\
 \hline
  9 & -0.25& 0.108 &0.096 &0.119  \\
  \hline
\end{tabular}
\caption{Comparison between numerical and calculated critical velocities using equations (\ref{uc}) and(\ref{rel})for $n=1,3,5,7,9$. }
\end{center}
\end{table}

 We note that the critical velocities are monotonically decreasing as $n$ increases because the barrier height deceases with increasing $n$ and at a very large $n$ the barrier disappears and the critical velocity approaches zero see figure 15.

\begin{figure}
\begin{center}
\includegraphics[angle=0, width=8cm]{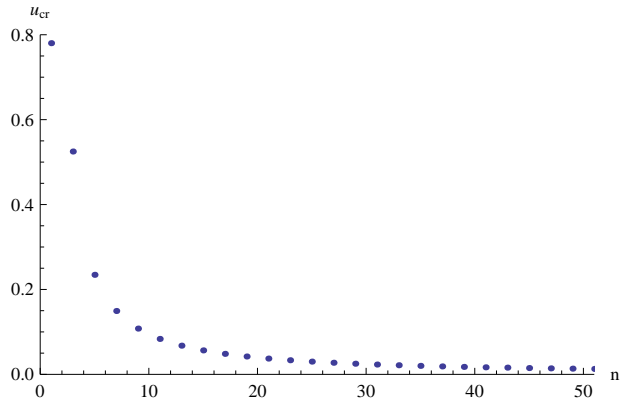}
\caption{$u_{cr}$ vs $n$ using equation (\ref{uc})}
\end{center}
\end{figure}






\section{\textbf{Conclusion}}

 In this paper we have considered a class of potentials using two parameters family ie. $a_{1}$ and $b_{1}$. These parameters control the shape, heights and widths of such potentials. This work gives rise to a class of potentials with $n=1,3,5,\dots$ that are themselves solutions to the field equation. This means that a static kink at presence of these potentials are explicitly known. In the asymptotic limit the static kink solution is also known and reduces to the familiar kink solution. We have only considered a particular parameter values ( $a_{1}=1$ ,$ b_{1}=\pm 0.25$)that generate various potentials which include pure barriers and wells, barriers with two side narrow wells, wells with two side humps, volcano barriers and double wells. We have examined the dynamics of $\varphi^{4}$ kinks with $n=1,3,5,\dots$ when they interact with these potentials. We have reproduced all relevant phenomena that are known such as trapping, back-reflection, and escape. With such potentials which are a mixture of barriers and wells,  an interesting dynamics have been seen and the trajectories, $x\left(t\right)$ have been plotted. As is the case in preceding work there is a critical velocity above which the soliton was able to escape the potential and below which the soliton is either reflect or trapped.

 We want to emphasize that the argument function $g\left(x\right)$ determines the scattering dynamics of solitons.

 We have then compared numerically calculated critical velocities to theoretical calculations using two different equations. There was a very good agreement between numerical and theoretical calculations. Also, Energy of these solitons at $x=0$ are also numerically and theoretically compared and we found that they have an exact agreement for $n=1$ and  agree to a very good extent for the others.

  In this paper we have chosen a particular soliton model namely $\varphi^{4}$ and have also selected a particular argument function of the field to generate kink solutions one is close to or at the obstruction and the other one faraway. However, we can apply the same work on other soliton models as well.

  Finally, An example of other possible arguments that satisfy all the constraints we set and generate kink solutions to $\varphi^{4}$ model and produce various smooth space dependent potentials is

 \begin{equation}
  g\left(x\right)=a_{1}x+\sum_{n=1}b_{n}\tanh^{n}\left(x\right), \qquad  n=1,2,3,4,5\dots.
 \end{equation}

  This type of argument function can be truncated to a certain $n$ and so we can have as many parameter family as we want from which we can generate various potentials. This will be left for future study.

 {\bf{Acknowledgement}}

 I would like to thank Professor Wojtek .J. Zakrzewski for his helpful comments.

\end{document}